\documentclass[aps,prb,showpacs,twocolumn,superscriptaddress]{revtex4}
\usepackage{amsmath}
\usepackage{amssymb}
\usepackage{graphicx}
\begin{document}
\title{Origin of the multiferroic spiral spin-order in the $R$MnO$_{3}$ perovskites}
\author{Shuai Dong}
\affiliation{Department of Physics and Astronomy, University of Tennessee, Knoxville, TN 37996, USA}
\affiliation{Materials Science and Technology Division, Oak Ridge National Laboratory, Oak Ridge, TN 32831, USA}
\affiliation{Nanjing National Laboratory of Microstructures, Nanjing University, Nanjing 210093, China}
\author{Rong Yu}
\author{Seiji Yunoki}
\affiliation{Computational Condensed Matter Physics Laboratory, RIKEN, Wako, Saitama 351¨C0198, Japan}
\author{J.-M. Liu}
\affiliation{Nanjing National Laboratory of Microstructures, Nanjing University, Nanjing 210093, China}
\affiliation{International Center for Materials Physics, Chinese Academy of Sciences, Shenyang 110016, China}
\author{Elbio Dagotto} 
\affiliation{Department of Physics and Astronomy, University of Tennessee, Knoxville, TN 37996, USA}
\affiliation{Materials Science and Technology Division, Oak Ridge National Laboratory, Oak Ridge, TN 32831, USA}
\date{\today}
\begin{abstract}
The origin of the spiral spin-order in perovskite multiferroic manganites $R$MnO$_{3}$ ($RE=$ Tb or Dy) is
here investigated using a two $e_{\rm g}$-orbitals double-exchange model. Our main result is that the experimentally 
observed spiral phase can be stabilized by introducing a relatively weak next-nearest-neighbor superexchange coupling ($\sim10\%$ 
of the nearest-neighbor superexchange). Moreover, the Jahn-Teller lattice distortion is also shown to be essential 
to obtain a realistic spiral period. Supporting our conclusions, the generic phase diagram of undoped perovskite 
manganites is obtained using Monte Carlo simulations, showing phase transitions from the A-type antiferromagnet, 
to the spiral phase, and finally to the E-type antiferromagnet, with decreasing size of the $R$ ions. These results are
qualitatively explained by the enhanced relative intensity of the superexchanges.
\end{abstract}
\maketitle

\section{Introduction}
Perovskite manganites, one of the main families of strongly correlated electronic materials, have drawn 
much attention since the discovery of  the colossal magnetoresistance in the last decade. The strong coupling 
between spin, charge, orbital, and lattice degrees of freedom gives rise to many competing phases in manganites, 
with rich physical properties. Theoretically, the physics of manganites appears to be qualitatively understood 
within the framework of the double-exchange (DE) model, including the superexchange (SE) between the $t_{\rm 2g}$ 
spins, and the Jahn-Teller (JT) interactions.\cite{Dagotto:Prp}

Recently, the multiferroic materials, in which the ferroelectric (FE) and magnetic orders coexist and are intimately coupled, 
have become a subject of much attention due to their technological relevance and fundamental 
science challenges.\cite{Eerenstein:Nat,Cheong:Nm} The discovery of multiferroicity in undoped manganites with 
small size $R$ (rare-earth) cations, e.g. TbMnO$_3$ and DyMnO$_3$, have made the family of manganites even more 
fascinating.\cite{Kimura:Nat,Goto:Prl} With decreasing temperature, these multiferroic materials $R$MnO$_3$ 
first transform from a paramagnetic (PM) state to a collinear spin sinusoidal incommensurate state at $T_{\rm N}$ ($\sim40-50$ K), 
and then to a spiral spin state with a locked period at $T_{\rm lock}$ ($\sim15-30$ K).\cite{Kenzelmann:Prl,Kimura:Prb05,Arima:Prl} 
$T_{\rm lock}$ is also the ferroelectric (FE) critical temperature ($T_{\rm c}$), suggesting a strong magnetoelectric 
interaction. In the collinear spin sinusoidal incommensurate state ($T_{\rm lock}=T_{\rm c}<T<T_{\rm N}$) 
of Tb$_{1-x}$Dy$_{x}$MnO$_3$, all spins approximately point into the $b$ direction and their amplitudes are modulated 
by the Mn positions in a sinusoidal form. The wavelength of this sinusoidal modulation, which is incommensurate to the 
lattice spacing, changes monotonously with decreasing temperature until $T_{\rm lock}$ is reached. Below $T_{\rm lock}$, 
the spins form a coplanar ($b-c$) spiral order, and its wavelength is nearly independent of temperature.\cite{Arima:Prl}

Such a spiral spin-order (SSO) driven improper ferroelectricity is also observed experimentally in other 
transition-metal oxides besides undoped manganites, such as Ni$_{3}$V$_{2}$O$_{8}$,\cite{Lawes:Prl} 
CuFeO$_{2}$,\cite{Kimura:Prb06} CuO,\cite{Kimura:Nm} and others.\cite{Kimura:Armr} However, the theoretical 
understanding of their properties is still in its early stages. For example, the origin of the FE polarization remains 
under much debate. Phenomenologically, the SSO breaks the spatial inversion symmetry and allows the emergence of a 
spontaneous polarization.\cite{Mostovoy:Prl06} Microscopically, two possible mechanisms for the FE polarization were 
proposed: (a) a pure electronic contribution driven by the spin-orbital coupling,\cite{Katsura:Prl,Jia:Prb} and  (b) 
a cation displacement driven by the Dzyaloshinskii-Moriya (DM) interaction.\cite{Sergienko:Prb,Li:Prb} Interestingly, 
both mechanisms give the same behavior: a FE polarization 
$\textbf{P}\propto-\textbf{e}_{i,j}\times(\textbf{S}_{i}\times\textbf{S}_{j})$, with $\textbf{e}_{i,j}$ being 
the unit vector connecting the nearest-neighbor (NN) spins $\textbf{S}_{i}$ and $\textbf{S}_{j}$. Although recent 
studies suggest the cation displacement to be the dominant one in manganites,\cite{Jia:Prb07,Xiang:Prl,Malashevich:Prl} 
further work is necessary to clarify the nature of the FE polarization in these materials.

In addition, it is important to remark that the origin of the SSO is also a puzzle. A direct but phenomenological 
route to generate a SSO phase is via the magnetic frustration between NN ferromagnetic (FM) and next-nearest-neighbor (NNN) 
antiferromagnetic (AFM) interactions,\cite{Cheong:Nm} e.g. via a $J_{1}$-$J_{2}$-$J_{3}$ model with classical 
spins, where $J_{1}$ is the NN superexchange while $J_{2}$ ($J_{3}$) is the NNN superexchange along the $b$ ($a$) 
directions.\cite{Kimura:Prb} In this model, a strong $J_{2}$ (compared with $J_{1}$) is required to generate a realistic 
short-wavelength spiral order. In fact, the minimum $J_{2}/|J_{1}|$ for the spiral phase is $0.5$ and to reach a 
realistic wavevector $q$, then $J_{2}/|J_{1}|$ should be larger, e.g. $0.8\sim1$.\cite{Kimura:Prb} 
However, in first-principles calculations,\cite{Xiang:Prl} the value of $J_{2}$ was found to be only $0.56|J_{1}|$,
which is not sufficient to induce the experimentally observed short-length ($L=6$) spiral order. 
Also, the coupling $J_{3}$ was found to be $0.375|J_{1}|$ and in addition it is AFM, opposite to the weak FM coupling ($0.01J_{1}$) 
used in the previous model study.\cite{Kimura:Prb} Furthermore, while providing interesting
information for the spiral regime,
it should be noted that the classical spin model is 
not suitable to explore in a single framework 
the many phases (A-AFM, spiral, E-AFM) found in $R$MnO$_3$ manganites, 
since it can not distinguish the E-AFM order from the 
$q=1/4$ spiral order (see Fig.~1) since they have the same energy. In summary, while 
the classical spin model with NNN magnetic 
frustration provides a natural starting point to describe phenomenologically 
the SSO phase, this simple model is 
not sufficient to deeply understand the microscopic origin of the SSO in perovskite manganites.

An alternative route to obtain the SSO phase is to incorporate the DM interaction ($\propto \textbf{S}_i\times \textbf{S}_j$) into 
the double-exchange framework.\cite{Sergienko:Prb,Li:Prb} However, the required intensity of the DM interaction 
is two orders of magnitude higher than expected.\cite{Sergienko:Prb} Thus, it is fair to express that the real 
driving force for SSO in manganites remains a mystery. Note that a more fundamental theory to explain the origin 
of the SSO phase should also explain the general phase diagram of $R$MnO$_{3}$: with decreasing $R$ size, 
the ground state of $R$MnO$_{3}$ changes from the A-type AFM (A) state, as in the case of LaMnO$_{3}$, to 
the spiral (S) state, as it occurs in TbMnO$_{3}$, and finally to the E-type AFM (E) phase, which is stabilized in HoMnO$_{3}$. 
This property is referred below as the ``A-S-E transition''.\cite{Goto:Prl,Zhou:Prl} For all these phases, 
the order of the Mn$^{3+}$ spins can be characterized by a particularly propagation vector ($0$, $q_{\rm Mn}$, $1$) 
(in the orthorhombic $Pbnm$ cell notation, see Fig.~1), indicating an AFM coupling along the $c$ axis, and FM coupling 
along the $a$ axis. $q_{\rm Mn}$ is $0$ for the A phase and $0.5$ for the E phase. 
However, for convenience when using a square lattice as in our simulations, $q$ in the following will be defined 
along the $x/y$ directions (see Fig.~1) and it equals half of the $q_{\rm Mn}$ used in experimental papers.

\begin{figure}
\includegraphics[width=240pt]{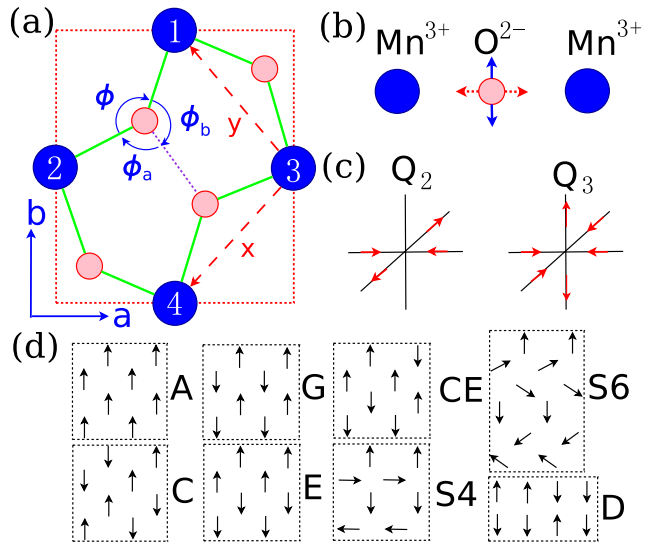}
\caption{(Color online) (a) Sketch of the crystal structure ($a-b$ plane) of $R$MnO$_{3}$. Two types of coordinate 
axes ($a-b$ and $x-y$) are shown. (b) Illustration  of the two kinds of distortions discussed in the text: 
GdFeO$_{3}$-type (oxygen moves perpendicular to the Mn-O-Mn bond) and Jahn-Teller type (oxygen moves along Mn-O-Mn bond). 
(c) Two Jahn-Teller distortion modes: $Q_{2}$ and $Q_{3}$. 
(d) Sketch of the spin patterns (in the $a-b$ plane) corresponding to seven ($S4$ and $S6$ belong to the same group) candidate phases considered here 
in the zero-temperature variational method. A, G, CE, C, and E are standard notations for well-known AFM phases in manganites. 
Here, the ferromagnetic chains in the C/CE/E-AFM phases are within the $a-b$ plane. D denotes 
the $\uparrow\uparrow\downarrow\downarrow$ dimer phase, that is considered in our studies although it has not been observed 
in real manganites.\cite{Aliga:Prb} In the spiral phases $SL$, the length period is denoted by $L$ (along $x$/$y$).
 The $L=4$ case has the same energy as the E-AFM phase in the classical spin model context.}
\end{figure}

\section{Model and Methods}
In this study, we will analyze the properties of the two $e_{\rm g}$-orbitals DE model for manganites defined on
a two-dimensional ($a$-$b$ plane) lattice to try to unveil the origin of the SSO phase in $R$MnO$_{3}$. 
We will start from a pure DE model, and then incorporate other interactions one-by-one to clarify their 
respective roles. The primary Hamiltonian, considering only the DE and NN SE interactions, reads as:
\begin{equation}
H_{\rm DE+SE} = - \sum_{<ij>}^{\alpha\beta} t_{\textbf{r}}^{\alpha\beta}\Omega_{ij} c_{i\alpha}^{\dagger} c_{j\beta} + J_{\rm AF} \sum_{<ij>}\textbf{S}_{i} \cdot \textbf{S}_{j},
\end{equation}
where the first term is the standard DE kinetic energy of the $e_{\rm g}$ electrons. 
The DE hopping amplitudes $t_{\textbf{r}}^{\alpha\beta}$ are orbital- and direction-dependent. In particular, 
$t_{x}^{1,1} = t_{y}^{1,1} = 3t_{x}^{2,2} = 3t_{y}^{2,2} = \frac{3}{4}t_{0}$, $t_{y}^{1,2} = t_{y}^{2,1} 
=-t_{x}^{1,2} = -t_{x}^{2,1} = \frac{\sqrt{3}}{4}t_{0}$ where the superscript $1$ ($2$) denotes the $e_{\rm g}$ 
orbital $d_{x^2-y^2}$ ($d_{3z^2-r^2}$), and $t_{0}$ ($0.2\sim0.3$ eV) is taken as the energy unit.\cite{Dagotto:Prp} 
The infinite Hund coupling used here generates a Berry phase 
$\Omega_{ij} = \cos(\theta_i/2)\cos(\theta_j/2) +\sin(\theta_i/2)\sin(\theta_j/2)\exp[-i(\varphi_i-\varphi_j)]$, 
where $\theta$ and $\varphi$ are the angles defining the $t_{\rm 2g}$ spins $\textbf{S}$ in spherical coordinates. 
The $e_{\rm g}$ kinetic (DE) energy is obtained by exactly diagonalizing the fermionic sector, using library subroutines,
once a $t_{\rm 2g}$ classical spins configuration is generated. The second term in the Hamiltonian is the usual AFM SE coupling between NN $t_{\rm 2g}$ spins.

The numerical methods used in this manuscript are of two varieties: the variational method employed
at zero temperature and the Monte Carlo (MC) simulation at finite temperature. For the case of the zero temperature variational method, 
the total energies (per site) of several candidate phases (for an infinite size lattice) are calculated 
to determine which is the most likely ground state. The candidate phases are sketched in Fig.~1, 
and almost all of the typical spin order patterns discussed in manganites have been included. 
In addition, spirals with wavevectors $q$ (from $0$ to $1/4$) are also taken into account. 
For the MC simulation, the classical $t_{\rm 2g}$ spins evolve in MC time following the standard Metropolis 
algorithm. The MC simulation is independent of the zero-temperature variational method,  namely the spin order obtained 
via MC simulations can be one of the candidate phases, but the outcome is unbiased and it is not limited to be
one of those phases. Due to the restriction of working on a finite lattice size, 
a large enough cluster should be used in the MC simulation to allow for other possible spin orders. 
Since the MC method used here has been discussed extensively in the manganite literature,
for more details regarding the Hamiltonian and the MC technique the readers are referred 
to Ref.~\onlinecite{Dagotto:Prp}

\section{Zero Temperature Results}
For the model studied here, the $e_{\rm g}$ electronic density is always $1$ when the chemical potential equals $0$. Thus, in the ``clean limit'' two-orbitals DE model without 
quenched disorder, the undoped $R$MnO$_{3}$ case is free from 
electronic phase separation tendencies, which usually occurs in doped cases.\cite{Dagotto:Prp} 
Therefore, it is reasonable to consider only homogeneous phases as candidates for the ground states. 
A very important result in this context is that using Eq.~1 
we have observed that the total energy of the spiral phase is always {\it higher} 
than the energy of both the A and the E phases, regardless of the value of $J_{\rm AF}$. 
In other words, the spiral phase can not be the ground state of Eq.~1, 
in agreement with previous studies of undoped manganites.\cite{Hotta:Prl} This model 
has to be modified for the
spiral state to become the ground state in some range of parameters.

\subsection{The role of the NNN superexchange}
In the rest of the paper, the spin frustrating effects of the NNN  $J_{2}$ coupling will be taken into 
consideration to solve this paradox. Here, $J_{2}$ arises from the super-exchange between NNN $3d$ spins. 
It should be noted that our model is conceptually different from those based on classical spins presented 
in previous publications.\cite{Cheong:Nm,Kimura:Prb} First, our NN SE is antiferromagnetic while it was
ferromagnetic in those previous efforts. Ferromagnetic tendencies do exist in our model but they originate 
in the DE mechanism of itinerant $e_{\rm g}$ electrons. Second, as shown in Fig.~1, considering the substantial distance 
between the NNN spins via the zigzag exchange path Mn-O-O-Mn, then $J_{2}$ should be weaker than $J_{\rm AF}$, 
in contrast to the robust $J_{2}$ used before.\cite{Cheong:Nm,Kimura:Prb} 
As indicated in Fig.~1, the GdFeO$_{3}$-type distortion of the oxygen octahedra, which reduces the NN 
Mn-O-Mn angle $\phi$, is crucial for $J_{2}$, because it shortens the distance between two oxygens and increases 
the Mn-O-O-Mn angles. Therefore, $J_{2}$ is enhanced by the GdFeO$_{3}$-type distortion.\cite{Kimura:Prb} 
Another important property of $J_{2}$ is its anisotropy due to $b>a$. For a crude estimation, if all Mn$^{3+}$ 
and O$^{2-}$ in Fig.~1 are simplified to be coplanar, it is easy to calculate the bond angles 
$\phi_{b}$ and $\phi_{a}$ from the experimental data available for $a$, $b$, and $\phi$. For the case of 
TbMnO$_{3}$, it was found that $b/a\approx1.1$  and $\phi\approx145.3^\circ$ suggesting 
$\phi_{b}\approx115.2^\circ$ and $\phi_{a}\approx99.5^\circ$.\cite{Goto:Prl,Kenzelmann:Prl} 
The larger bond angle leads to a stronger NNN spin coupling through the Mn(1)-O-O-Mn(4) exchange path. 
Considering the angle-dependence of the super-exchange,\cite{Zhou:Prb} the NNN SE $J_{\rm 2b}$ coupling 
between Mn(1) and Mn(4) can be between $1.8$ to $2.2$ times the value of the $J_{\rm 2a}$ coupling 
between Mn(2) and Mn(3). In addition, the $e_{\rm g}$ orbital-order can also induce this 
anisotropy.\cite{Kimura:Prb} Thus, the new term in the Hamiltonian incorporating the NNN SE coupling is
\begin{equation}
H_{\rm J2}=\sum_{[ij]}J_{2\gamma}\textbf{S}_{i}\cdot\textbf{S}_{j},
\end{equation}
where $\gamma$ runs over $a$ and $b$,  and [{\it ij}] denotes NNN sites. 
Considering together Eqs.~1 and 2, a zero-temperature phase diagram can be obtained 
by comparing the total energies of the candidate phases discussed in Fig.~1(d). As shown in Fig.~2(a), 
it is remarkable that now the spiral phase becomes the ground state for values of $J_{\rm 2b}$ as small as $0.017$.

\begin{figure}
\includegraphics[width=260pt]{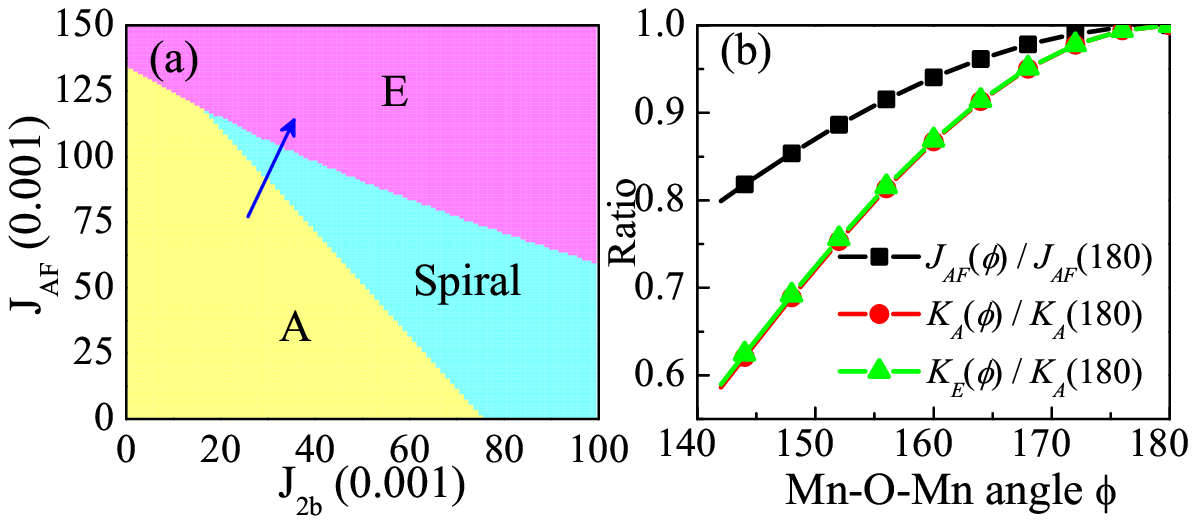}
\includegraphics[width=240pt]{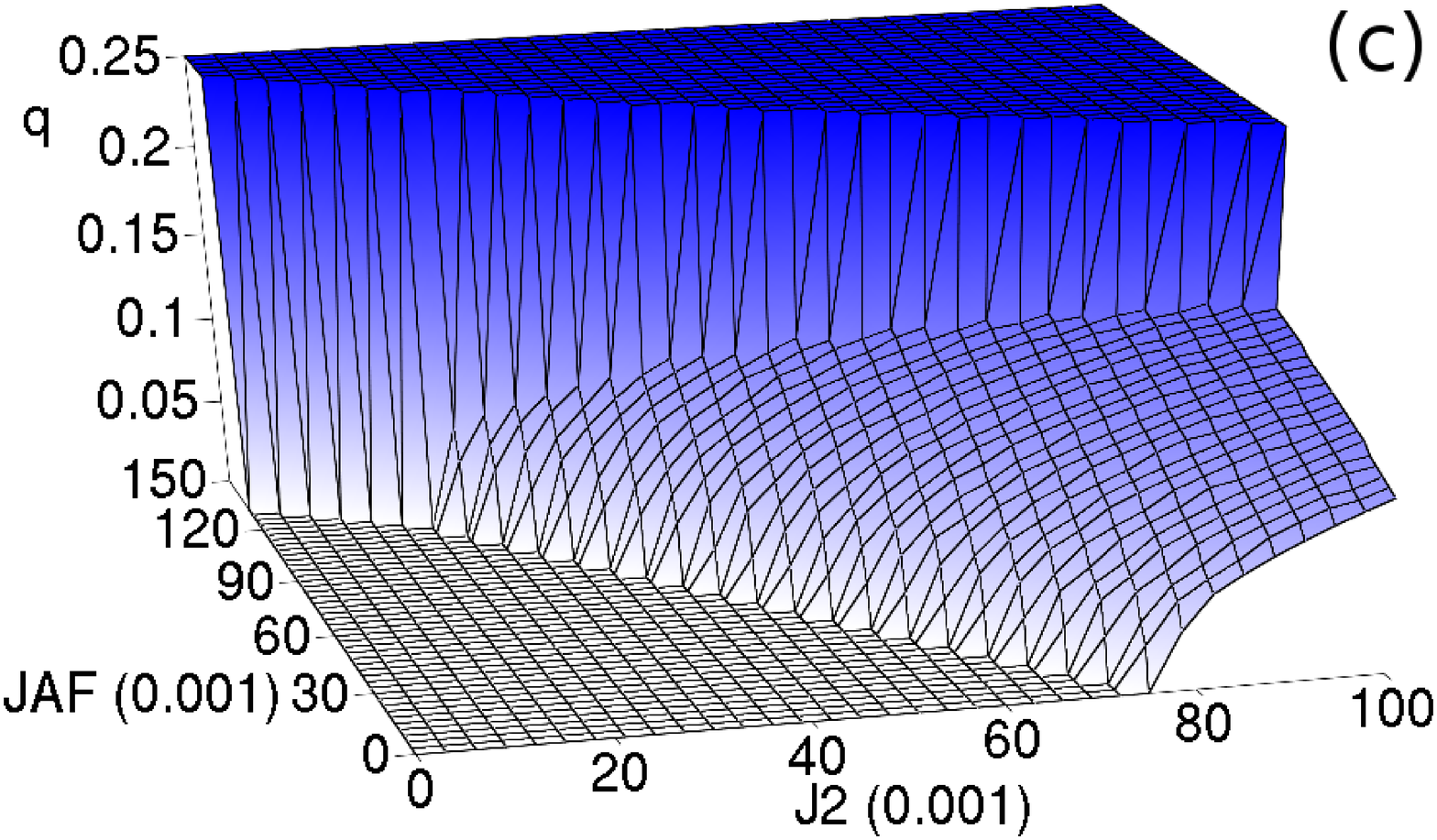}
\caption{(Color online) (a) Zero-temperature phase diagram of the two-orbitals DE model for $R$MnO$_{3}$. 
The possible path for the A-S-E phase transition is indicated by the arrow. The phase diagram is independent of $J_{\rm 2a}$, 
as long as $J_{\rm 2a}<J_{\rm 2b}$. (b) $J_{\rm AF}$, $K_{\rm A}$ and $K_{\rm E}$, as a function of the Mn-O-Mn angle, 
normalized to their values at $\phi=180^\circ$. (c) Wavevector number $q$ for (a). Here, $q$ is defined along 
the $x/y$ direction and equals half of the $q_{\rm Mn}$ used in experimental papers.}
\end{figure}

Considering the GdFeO$_{3}$-type distortion, we calculated the $\phi$-dependent 
$J_{\rm AF}$ (using the relation $J_{\rm AF}\propto\sin^4(\phi/2)$ described in Ref.~\onlinecite{Zhou:Prb})
and also the DE kinetic energy (using 
$t^{1,1}\propto\cos^3\phi$, $t^{1,2}$ and $t^{2,1}\propto\cos^2\phi$, and $t^{2,2}\propto\cos\phi$, 
as in Ref.~\onlinecite{Sergienko:Prl})
of the A and E phases. The latter are denoted by $K_{\rm A}$ and $K_{\rm E}$, respectively, and are shown in Fig.~2(b).
Both the DE kinetic energy and $J_{\rm AF}$ decrease with decreasing $\phi$. 
The curves for $K_{\rm A}$ and $K_{\rm E}$ are almost identical and lower than the $J_{\rm AF}$ curve. 
Therefore, the GdFeO$_{3}$-type distortion increases $J_{\rm AF}$ relative to the DE kinetic energy. 
Noting that $J_{\rm 2}$ is enhanced by the GdFeO$_{3}$-type distortion, then the total effect of this distortion 
is the enhancement of both the NN and NNN SE couplings, particularly the latter. Therefore, the A-S-E phase 
transition with decreasing $R$ size can now be easily understood, and the arrow in Fig.~2(a) 
indicates a possible path that is qualitatively in agreement with experiments. 
Below, we will still use the ideal case $\phi=180^\circ$, as in most previous studies, since the GdFeO$_{3}$-type 
distortion effect can be accounted for by increasing the relative values of $J_{\rm AF}/t_0$ and $J_{\rm 2}/t_0$ (see Fig.~2b).

As shown in Fig.~2(c), at the A to S transition the wavevector $q$ changes continuously from $0$ to a finite value. 
However, between the spiral and E phases, not only $q$ now changes discontinuously, but also their spin patterns 
are incompatible, namely the E-phase is not a special case of the spiral phase. These two different phase boundaries 
may cause two distinct behaviors when $R$ cations substitutions occur: (a) if Tb is substituted by an 
A-phase $R$ (e.g. Gd), $q$ will be reduced continuously; (b) if Dy is substituted by an E-phase $R$ (e.g. Ho), phase separation at some concentrations may be induced.

\subsection{The role of the Jahn-Teller couplings}
 In spite of the success in describing the SSO phase in the previous section, we have 
observed that in Fig.~2(c) the wavevector $q$ is still smaller than the experimental value. 
For example, to obtain the $q$ for TbMnO$_{3}$ ($0.14$ along the $x$/$y$ direction), the minimum $J_{\rm 2b}$ 
is $0.073$ which corresponds to  $J_{\rm AF}$=$0.074$. This $J_{\rm 2b}/J_{\rm AF}$ ratio is too large for real manganites, 
as mentioned before in this text. Moreover, the largest $q$ in Fig.~2 can not reach the experimental 
value for DyMnO$_{3}$ ($q$=$0.19$). Even if the GdFeO$_3$-type distortion is taken into 
account ($\phi$$\approx$$144.7^\circ$ for DyMnO$_{3}$),\cite{Goto:Prl} $q$ can not be increased further.

Interestingly, we have observed that realistic values for $q$ can be stabilized if the Jahn-Teller distortion (Fig.~1(b)) 
is incorporated. This JT distortion can be characterized by two modes: $Q_{2}$ and $Q_{3}$ (Fig.~1(c)), 
which couple with the $e_{\rm g}$ electrons via:
\begin{equation}
H_{\rm JT}=\lambda\sum_{i}[Q_{2,i}\tau_{x,i}+Q_{3,i}\tau_{z,i}],
\end{equation}
where $\lambda$ is the spin-phonon coupling coefficient and $\tau$ is the orbital pseudospin operator, 
given by $\tau_{x}=c_{a}^{\dagger}c_{b}+c_{b}^{\dagger}c_{a}$ and $\tau_{z}=c_{a}^{\dagger}c_{a}-c_{b}^{\dagger}c_{b}$.\cite{Dagotto:Prp} 
For all $R$MnO$_{3}$ at low temperatures, $|Q_{2}|$ and $Q_{3}$ are uniform with $|Q_{2}|\approx-\sqrt{3}Q_{3}$.
The sign of $Q_{2}$ is staggered, which gives rise to the well-known 
staggered $d_{\rm 3x^2-r^2}$ and $d_{\rm 3y^2-r^2}$ orbital order.\cite{Zhou:Prb} 
Therefore, Eq.~3 can be scaled using $\lambda|Q_{2}|$ as the only parameter.

\begin{figure}
\includegraphics[width=260pt]{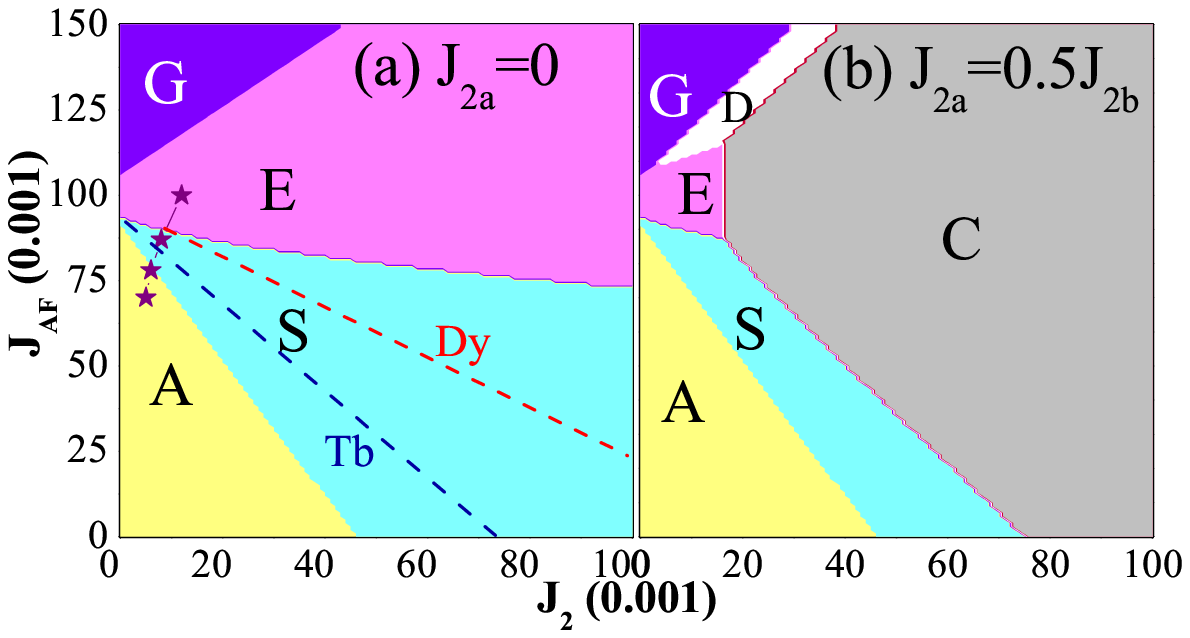}
\includegraphics[width=240pt]{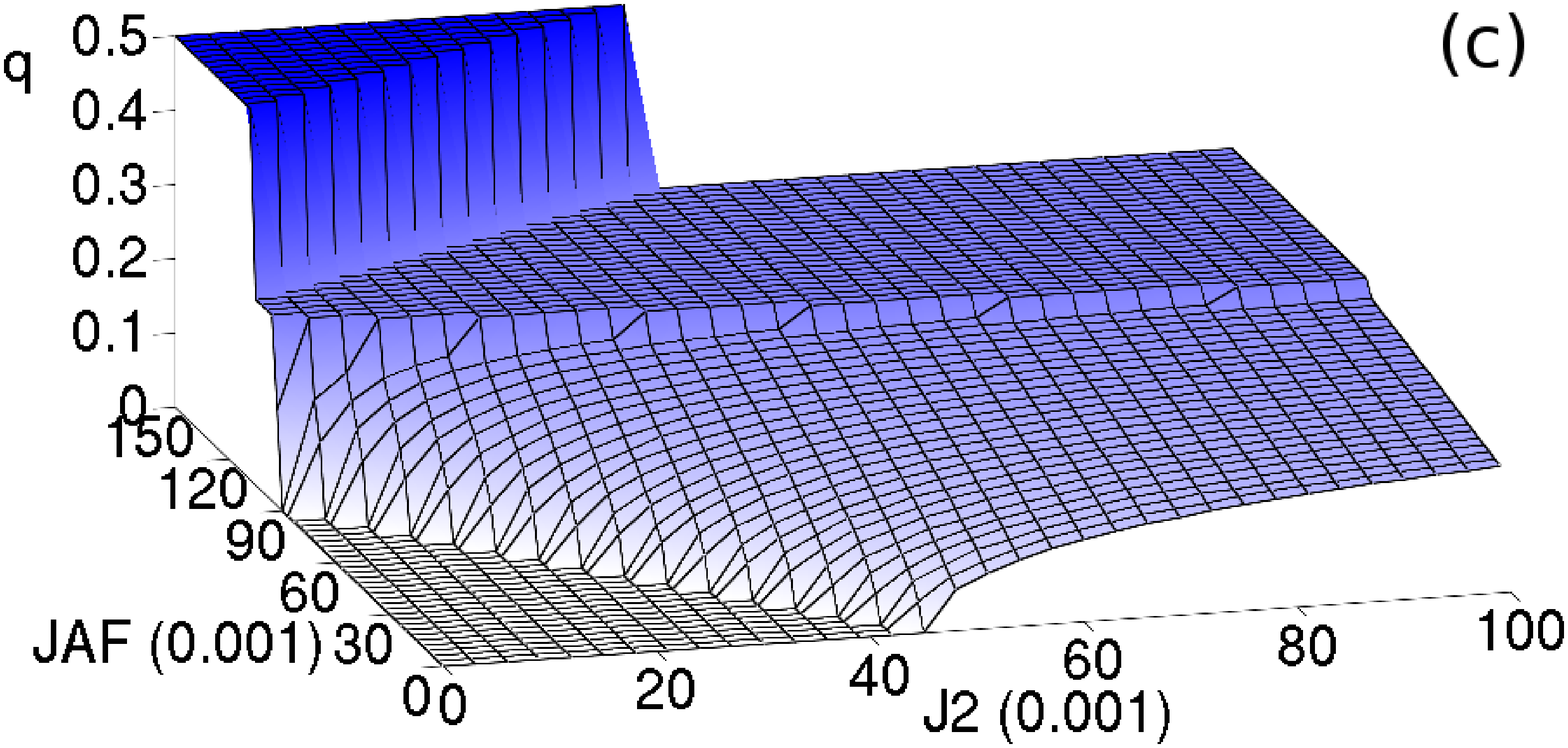}
\caption{(Color online) (a) Zero-temperature phase diagram of the two-orbitals DE model with JT distortions 
($\lambda|Q_{2}|$=$1.5$) for $R$MnO$_{3}$. The possible values of $J_{\rm AF}$-$J_{\rm 2b}$ for realistic $q$s in TbMnO$_{3}$ 
and DyMnO$_{3}$ are shown with dashed lines. The four asterisks represent the set of couplings (1st and 2nd for A, 3rd for S, 
and 4th for E) that are studied via the MC techniques. (b) Same as (a) except for a finite $J_{\rm 2a}$=$0.5J_{\rm 2b}$. 
The four asterisks in (a) remain within the A-S-E region here. (c) Wave vector number $q$ for (a).}
\end{figure}

Considering together  Eqs.~1, 2 and 3, we recalculated the zero-temperature phase diagram. A typical result 
with $\lambda|Q_{2}|=1.5$ is shown in Fig.~3. Compared with Fig.~2(a), now the spiral phase region in Fig.~3(a) is enlarged. 
This spiral phase can even be stable without $J_{\rm 2b}$, although in a very narrow $J_{\rm AF}$ region. 
In Fig.~3(c), the wavevector $q$ of the spiral phase is increased as a whole, compared with those in Fig.~2(c). 
Therefore, the $q$ for TbMnO$_{3}$ and DyMnO$_{3}$ can be obtained with a small $J_{\rm 2b}$, e.g. the minimum 
$J_{\rm 2b}$ is now $0.002$ ($J_{\rm AF}$=$0.091$) for TbMnO$_{3}$, and $0.008$ ($J_{\rm AF}$=$0.09$) for DyMnO$_{3}$, respectively. 
In other words, with the JT distortion incorporated, 
a rather weak NNN SE ($<10\%$ of NN SE) is enough to generate the realistic SSO in $R$MnO$_{3}$. 
In contrast to the case without JT distortions, $J_{\rm 2a}$ also plays a prominent role here, although it can not 
change the phase boundary between the A-S-E phases. Moreover, Fig.~3(b) shows that a robust $J_{\rm 2a}$ even 
induces C and D phases in the large $J_{\rm 2b}$ region, implying that both $J_{\rm 2b}$ and $J_{\rm 2a}$ in real manganites 
should be weak since none of these two phases have been observed experimentally in bulk $R$MnO$_{3}$.

Besides providing realistic values for $q$, the JT distortion can contribute to the insulating nature of 
$R$MnO$_{3}$, which is crucial for the FE polarization. For instance, with $\lambda|Q_{2}|=1.5$, the calculated gap (not shown)
is $2.59$ ($\sim0.52-0.78$ eV) for the spiral phase at the $q$ of TbMnO$_{3}$. 
The agreement with experimental results for the gap 
($0.489$ eV) suggests that the chosen value for $\lambda|Q_{2}|$ is reasonable.\cite{Cui:Ssc}

To understand the enhancement of $q$ by the JT effect, let us focus on the A to S transition. 
We consider $\varphi_i=0$ and $\theta_i=2\pi\textbf{q}\cdot\textbf{r}_i$ for the $t_{\rm 2g}$ spins. After diagonalizing 
the fermionic operators in Eqs.~1-3, the ground state energy per site is 
$E= K+2J_{\rm AF}\cos(\Delta\theta)+J_{\rm 2b}\cos(2\Delta\theta)+J_{\rm 2a}$, where $\Delta\theta=2\pi q$ is 
the angle between NN spins, and the fermionic sector energy is $K\approx K_{\rm A}\cos(\frac{\Delta\theta}{2})$. 
Expanding $E$ around $\Delta\theta=0$, we obtain 
$E\approx E_{0}-(\frac{K_{\rm A}}{8}+J_{\rm AF}+2J_{\rm 2b})\Delta\theta^2+(\frac{K_{\rm A}}{384}+\frac{J_{\rm AF}}{12}+\frac{2J_{\rm 2b}}{3}) \Delta\theta^4 +O(\Delta\theta^6)$, 
where $E_{0}$ is a constant and $\Delta\theta$ (hence $q$) becomes an order parameter. 
The A-S phase boundary is determined by $\frac{K_{\rm A}}{8}+J_{\rm AF}+2J_{\rm 2b}=0$, 
which agrees with the numerical results shown in Figs.~2(a) and 3(a-b). From these considerations, 
it is clear that the competition between $K_{\rm A}$, $J_{\rm AF}$, and $J_{\rm 2b}$ is crucial 
to obtain the spiral phase using realistic interactions, distinguishing our model from the previously used 
effective $J_{1}-J_{2}$ spin models. Furthermore, the reduction of $-K_{\rm A}$ by the JT distortion 
lowers the threshold of ($J_{\rm AF}$, $J_{\rm 2b}$) for a spiral phase which, as a consequence, leads to an enhancement of $q$.

\section{Finite Temperature Results}
\begin{figure}[b]
\includegraphics[width=260pt]{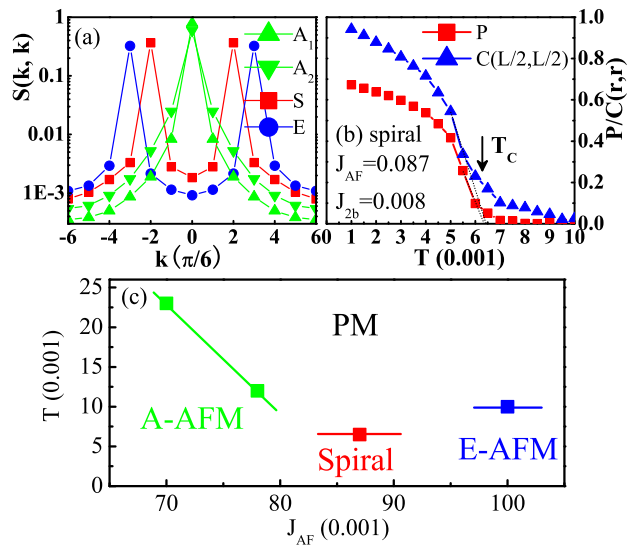}
\caption{(Color online) (a) Spin structure factor of the four MC simulations at $T=0.004$, corresponding to the four sets
of couplings described in the text. All characteristic peaks are prominent, suggesting stable spin orders. 
(b) Temperature-dependent polarization $P$ and $C(\frac{L}{2},\frac{L}{2})$ corresponding to the spiral phase 
with $q=1/6$. All values are normalized to their saturation values. (c) Sketch of the finite-temperature phase diagram. 
The $T_{\rm N}$'s for the A and E phases are determined from their $C(\frac{L}{2},\frac{L}{2})$-temperature curves. 
The lines connecting $T_{N}$ ($T_{c}$) are just to guide the eyes.}
\end{figure}

The above described phase diagrams were obtained at zero-temperature comparing the energies of several 
candidate phases. 
In principle, this procedure can not rule out the possibility of other unknown phases becoming stable. Therefore, 
it is necessary to check the results using MC techniques. For simplicity, the spins are restricted to be in an 
easy-plane for the magnetic moments, namely we use the $X-Y$ model instead of the Heisenberg model for the $t_{\rm 2g}$ spins. 
This simplification, which avoids the introduction of extra adjustable parameters to generate a 
magnetic anisotropy, is reasonable from the point of view of the physics of these materials since 
all spins are coplanar for all known low-temperature phases in $R$MnO$_{3}$. The JT distortions are fixed as in the previously 
described zero-temperature case. ${\bf P}$ is obtained using the model-free equation 
$-\textbf{e}_{i,j}\times(\textbf{S}_{i}\times\textbf{S}_{j})$. 
We use a 2-dimensional (2D) $L\times L$ ($L=12$) lattice with periodic boundary conditions and 
this computational study is carried out only at four sets of
couplings ($J_{2b}$, $J_{\rm AF}$), which are denoted by asterisks in Fig.~3(a). The very time consuming MC runs were typically carried out with 6000 steps for thermalization and 4000 steps for measurements.
Considering that all the physics of the four coupling sets (asterisks) is not much affected by a weak $J_{\rm 2a}$ 
(see Figs.~3(a) and (b)), here $J_{\rm 2a}$ is set to be zero for simplicity. 
The $q$ for the 3rd asterisk should be the commensurate value $1/6$ 
(suitable for Tb$_{0.41}$Dy$_{0.59}$MnO$_{3}$) to match the lattice size.\cite{Arima:Prl}

For all four pairs of couplings, the presence of sharp characteristic peaks in the spin structure factor 
of our MC results (Fig.~4(a)) confirms the stability of the candidate phases at low temperature. 
In addition, Fig.~4(b) shows that the $T_{\rm c}$ for the FE transition in the spiral phase is found 
to be $0.0065$ ($\sim16-24$ K), agreeing with the experimental value $22$ K.\cite{Arima:Prl} 
It should be noted that although the Mermin-Wagner theorem forbids a finite $T_{\rm c}$ (for an infinite lattice size) in 
the 2D $X-Y$ model, extensive MC studies on double-exchange models in the past decade have suggested that the 
characteristic $T^*$ in 2D small size lattices where the order develops (namely when the correlation length becomes comparable
to the lattice size) can still be comparable with the experimental critical temperatures, at least qualitatively. In a 2D regime of very large correlation lengths, a weak coupling in the third
direction rapidly establishes long-range order with $T^*$ as the critical temperature. 
The real-space correlation between spins at the maximum distance in our lattice $C(\frac{L}{2},\frac{L}{2})$ 
can also be used to provide estimations of the critical transition temperature.

Based on these considerations, here a finite-temperature phase diagram is sketched (Fig.~4(c)), 
guided by the four coupling sets of Fig.~3(a). It should be noted that in principle the collinear spin 
sinusoidal incommensurate state at $T_{\rm c}<T<T_{\rm N}$  is not possible in our study, because in our model 
the spin amplitude $|S|$ is a constant and the period of the phases should be commensurate with $L$. Despite 
this finite-size effect problem, our phase diagram nevertheless resembles at least qualitatively the real one, and 
moreover our rough estimations for the $T_{\rm N}$ of the A and E phases also agree well with experimental results.\cite{Goto:Prl}

\section{Conclusion}
In conclusion, here we have provided a microscopic description of
multiferroic $R$MnO$_{3}$ perovskites that explains the existence of the spiral spin-order
in these compounds. The experimentally observed SSO and FE transition can be 
obtained by incorporating a weak NNN superexchange interaction and a Jahn-Teller 
distortion into the standard two-orbitals DE model for
manganites. Several aspects of the experimentally known A-S-E phase 
transition with decreasing $R$ size 
are well reproduced by including the GdFeO$_{3}$-type distortion in our study.
Note that this weak NNN SE interaction ($<10\%$ NN superexchange), 
while shown here to be crucial in the context of the manganite 
multiferroics, it does not alter the previous large body of investigations and 
conclusions reached via similar MC simulations for undoped and doped LaMnO$_3$,
since in that case the extra NNN SE couplings can be neglected.

\section{Acknowledgment}
We thank N. Furukawa and K. Yamauchi for helpful discussions, and T. Kimura, S. Ishihara and T. Hotta for useful comments. This work was supported by the NSF (Grant No. DMR-0706020) and the Division of Materials Science and Engineering, U.S. DOE, under contract with UT-Battelle, LLC. S.D. and J.M.L. were supported by the National Key Projects for Basic Research of China (Grant No. 2006CB921802, 2009CB929501) and Natural Science Foundation of China (Grant No. 50832002). S.D. was also supported by the China Scholarship Council and the Scientific Research Foundation of Graduate School of Nanjing University.

\end{document}